\documentstyle[11pt,ysc,twoside,epsf]{article}
\def\edcomment#1{\iffalse\marginpar{\raggedright\sl#1\/}\else\relax\fi}
\reversemarginpar

\begin{document}
\title{An Interactive Program for Correlative Studies of Solar Energetic Particle Events}
\author{Amjad Al-Sawad, Jarmo Torsti}
\affil{Department of Physics and the V\"{a}is\"{a}l\"{a} Institute
for Space Physics and Astronomy, Turku University, FIN 20014 Turku,
Finland}
\author{Hannu Hoffr\'{e}n, Kari Lehtom\"{a}ki}
\affil{Savonia University of Applied Sciences, Engineering, Kuopio,
Finland}

\begin{abstract}
We have developed an interactive program which shows the solar
energetic particle (SEP) intensity-time profile as observed by
SOHO/ERNE, simultaneously with the associated coronal mass ejection
in optical imaging movies taken by LASCO coronagraph, soft X-ray by
YOHKOH, ultraviolet by EIT, DH radio emission by WAVE/Wind, and the
H$\alpha$ location for the solar flare and spectral radio emission
from the journal of geophysical data. The whole set of data will
provide increased scientific knowledge on the solar energetic
particle events and the solar phenomena associated with them,
because in this program one can see easily the temporal associations
of each phenomenon during the evolution of the particle intensity.
The (SEP) intensity-time profile will give a clear view to detect
the velocity dispersion in the events, if it exists. The ERNE data
are commented in order to follow of phenomena associated with
changes of the intensity-time profiles. We introduce this set of
data as an index for the ERNE/SOHO solar energetic particle events.
The interactive program is applied for statistical, correlative
study of SEP events observed on board SOHO.
\end{abstract}

\section{Introduction}

Solar energetic particle (SEP) events are one of the most
interesting phenomena in solar physics, which has been widely
observed near the Earth with energy ranges varying from some
keV/nucl to some GeV and they might have different sources such as
solar flare in the low corona, coronal shock and interplanetary
shocks driven by CMEs. Mason et al. (1999) suggested that high
energy and low-energy particles may result from different seed
populations and acceleration mechanisms. The sources of the SEP
events have been studied in the last two decades in details, but
still there are lots of arguments about the the main accelerators
for the energetic particles in different levels of energy. Different
classes of events may produce or accelerate different ranges of
energies and intensities. The intensity-time profile and the
associated solar phenomenon and composition studies are probably the
main sources for such studies.

In the 1980's the solar flares and coronal mass ejection CME have
been considered the main source for the SEP events. On the basis of
SEP events signature in soft X-ray Cane et al. (1986) found out two
classes: 1) impulsive events, which have high electron-to-proton
ratio, and are never associated with interplanetary shocks, but are
associated to impulsive flares that occur low in the corona, and 2)
gradual events, which can accelerate much higher proton energies and
are well associated with coronal and interplanetary shocks, and
occur high in the corona in extended regions.

To study the association of solar flares to the SEP events one has
to follow all the data related to soft X-ray emission and H$\alpha$
line absorption. The GOES series of satellites provide observations
through the {\it Solar Geophysical Data}. On the other hand, the
YOHKOH satellite provides continuous observation for the Sun in soft
X-ray emission. Coronal Mass Ejections (CMEs) have been shown to be
associated with Long Duration Events (LDEs) (Sheeley et al., 1975;
Kahler, 1977; and Sheeley et al., 1983), and with interplanetary
shocks (Sheeley et al., 1983, 1985).

The role of CMEs and solar flares in accelerating the SEP is not
clear yet. Gosling (1993), in his solar flare myth paper has moved
the central role of SEP from flares to CMEs. There are different
ideas about the role of coronal and interplanetary shocks. On the
one hand, Reames (1999) suggested that most intense SEP events, with
particles of highest energies, are produced by acceleration at
collisionless shock wave driven by CMEs. But on the other hand,
Kallenrode (1996) suggested that a CME-driven shock may not itself
accelerate significant numbers of particles out of the ambient solar
wind to high energies, but it can confine and re-accelerate
particles which were initially accelerated close to the Sun. Between
the two arguments, some recent studies indicate that SEPs may be
produced also on the global corona scale between the impulsive flare
and the interplanetary shock (Kocharov et al.,1999; Laitinen et
al.,2000; Klein \& Trottet, 2001). The LASCO coronagraph on board of
({\it SOHO}) is a rather good tool for getting available scientific
data concerning the CMEs. Thus, to study SEP events we have to
follow carefully each changing in the intensity-time profile and
find the link with the associated eruptions in the Sun.

Associations of radio emission with SEP events are very important to
determine characteristics of eruptions and acceleration mechanisms.
Kahler (1982) concludes that the occurrence of a type IV burst
"appears to be a requirement for most proton flares" at energies
$>$20 MeV. Type IV bursts are often, but not always, accompanied by
type II bursts, which reveal the passage of a large scale shock wave
through the corona (Klein \& Trottet, 2001).

Type II radio bursts are believed to be produced by shocks
propagating through the solar corona and interplanetary medium
(e.g., Riener et al., 2000). It is generally accepted that metric
type II (slow-drift) solar radio bursts are manifestations of
coronal shock waves caused by disturbances moving outward through
the solar atmosphere with speeds of several hundred kilometers per
second (Nelson \& Melrose, 1985). The existence of fast mode shocks
in the corona is strongly supported by the observation of rapidly
drifting radio bursts and their association with fast ($\geq$400
km/s) CMEs (e.g., Kahler, 1992).

There has been a long-standing controversy about the relation
between metric type II bursts, flares, CMEs and IP shocks (Chao,
1974, 1984; Wagner \& McQueen, 1983; Gosling, 1993; Gosling \&
Hundhausen, 1995; Svestka, 1995; Dryer, 1996; Gopalswamy et al.,
1998; Cliver et al., 1999). Using the data from November 1994 to
June 1998, Gopalswamy et al. (1998) reported that 93 metric type II
bursts did not have interplanetary signatures. On the other hand,
Cliver et al. (1999) insisted that metric type II, EIT wave, and
decimetric-hectometric type II bursts are driven by fast CMEs. The
recently observed EIT waves are found to be associated with metric
type II bursts (Klassen et al., 1997; Gopalswamy et al., 2000) in
their speeds and positions as well as with CMEs (Thompson et al.,
2000).

A distinction is generally made between type II radio bursts
observed at decimetric-metric wavelengths, referred to as coronal
type II bursts, and hectometric-kilometric wavelengths, referred to
as IP type II radio bursts. While IP type II bursts are usually
ascribed to bow shock waves driven ahead of a CME piston (Kahler
1992), the proposed origins for coronal type II bursts are still
debated. Some suggest that these are CME driven, like interplanetary
shocks (Cliver, Webb \& Howard, 1999). However, coronal type II
bursts are known to have a close temporal association with solar
flares (Swarup, Stone \& Maxwell, 1960; Dodge, 1975; Cane \& Reames,
1988).

Recently, Reiner et al. (2001) suggested that the harmonic component
of metric type IIs can be possibly related with
decimetric-hectometric type IIs. Also, Leblanc et al. (2001) argued
from 10 type II bursts that the shock waves may be driven by the
CMEs all the way from $\sim1R_{\bigodot}$ to 1 AU. In addition, they
admitted for some events that the evidence available cannot exclude
the hypothesis that the shock is a blast wave from the flare to 1 AU
(Smart \& Shea, 1985). Cho et al. (2003) reported that CMEs and
flares are initiated nearly simultaneously, at least for type II
associated events, based upon the temporal relationship obtained
from data from May 1998 to December 2000. Recently, it has been
suggested that CMEs and flares (metric type II) are initiated nearly
simultaneously (e.g., Zhang et al., 2001; Neupert et al, 2001; Moon
et al., 2002; Cho et al., 2003; Shanmugaraju et al., 2003).

We have provided all the above important data running
simultaneously with the intensity-time profile provided by
ERNE/SOHO to establish new view for studying the SEP events and to
achieve the result for such studies much more easily.

\section{The Data}

We have used intensity-time profile for proton and Helium particles
provided by the two ERNE's detectors, High Energy Detector (HED) and
Low Energy Detector (LED) (Torsti et al., 1997). We used the linear
intensity-time profile and compare the closest eruption on the Sun
to the first arrived protons on the high energy channels. The
SOHO/LASCO catalog at
$http://cdaw.gsfc.nasa.gov/CME_list/UNIVERSAL/$ has been used to
determine the closest CMEs to those events. For each event we
suggest one or more CMEs seen by Lasco and make the movies run
simultaneously with the intensity-time profile. We also make all the
features (speed, acceleration, central position angle and angular
width of the associated CMEs), available in the comments. Then we
repeat the same steps with the Yohkoh movies, to provide the
information about the solar flare through the observations of the
Soft X-ray Telescope SXT onboard Yohkoh. The EIT movies are also
provided to detect the EIT wave onset and give more information
about the associated solar flare. From the ({\it Solar Geophysical
Data}) we use the data concerning the metric radio emission
associated to the event. The decimetric-hectometric radio emission
data are provided through {\it WAVES/Wind}.

\section{The Program}

\begin{figure}
\plotone{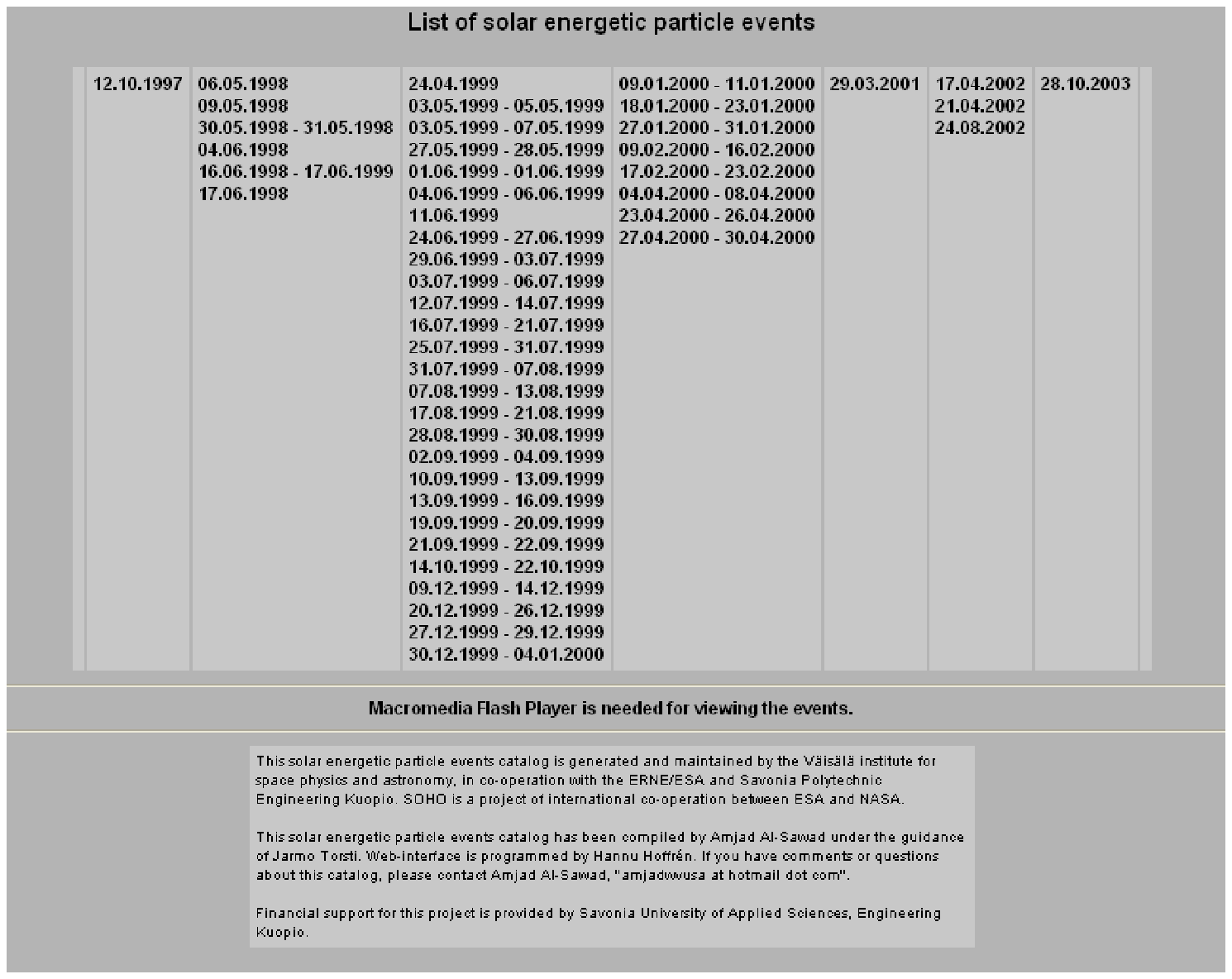} \caption{The interface of the interactive
program}
\end{figure}

First step in studying the SEP events through the interactive
program is to start with the catalogue interface \footnote{The
Interactive program is temporarily located at
http://www.it.savonia-amk.fi/hhoffren/soho/index.php}. Fig. 1 shows
the interface of the program. The time starts with the year that
{\it SOHO} started to operate and continues as long as it stays
operating. For an exact event in mind, one can just select the month
in the year that this event has occur. Look through that list for
chosen event and then click on it. Another interface will open Fig.
2. The chosen energy channels for that event will appear on the
upper left side of the page. The selection is based on the
fluctuation of the intensity-time profile. We normally choose the
less fluctuating energy channels with 10 minutes resolution. One can
choose the energy channel that he likes to study and run it, or he
can choose as many as he like and run them simultaneously. Each
energy channel has different color. In Fig. 2 we have chosen six
energy channels, three from protons and three from helium. The
vertical line, which moves with the mouse in the interface shows the
time at the position. On the same time the reading for the intensity
will appear in the channel list in the top left.

On the upper right side of the interface there is a button. Which
hides movies of Lasco, EIT and Yohkoh. Below the movies another
button, which shows the comments, metric and D-H radio emission.

\begin{figure}
\plotone{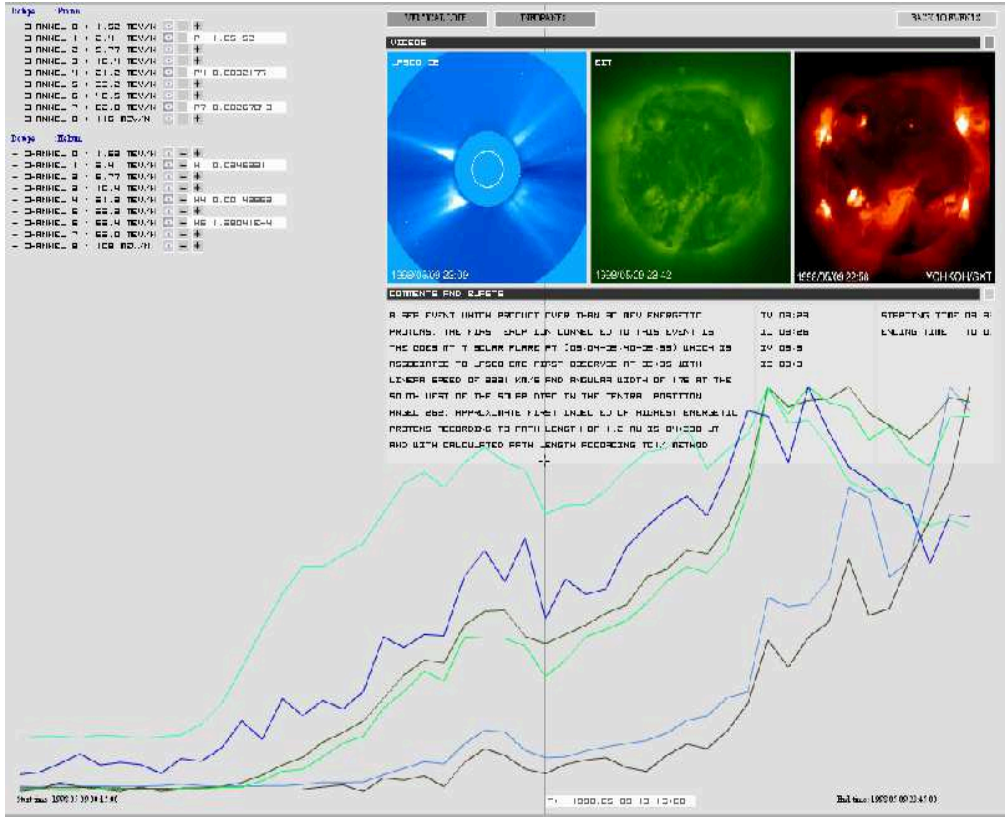} \caption{Demonstration for one of the analyzed
events, the event of 9th of May 1998}
\end{figure}

Normally we add our own comments to each event. The comments include
the suggested eruption at the Sun, which is thought to be the source
of the event. In some events there might be multiple eruptions.
Those events we denote as Multi-Eruption Solar Energetic Particle
Events (MESEP). An attempt to a statistical study for such events
are being worked out. Work is underway to mark each event as a high
energy or low energy event, depending on the highest energies
produced by that event. The information about the associated metric
and D-H radio emission were written according to the timing from the
start of the event.

\section{Measurements}
The program is using the linear plotting for the intensity-time
profile. This makes it easy to recognize the onset times of the
events which shows clear velocity dispersion, that is, a higher
energy channel starts to rise earlier than a lower energy one (see
Fig. 2). When you have such event with velocity dispersion it means
that it is mostly related to an eruption at the Sun. Thus we need to
calculate the injection time of the event and compare it to the
eruptions suggested by Lasco, EIT and Yohkoh.

The calculation for the injection time of the particles at the Sun
can be achieved through two ways. The first way, similar to Torsti
et al. (1999), is to consider that the non-scattered particles of
the chosen energies are propagating along the Archimedean field line
of nominal length 1.2 AU and calculate the flight time of those
particles by using the particle energy. Then subtract the flight
time from the observed time on ERNE and add 500 sec to get the
observed time fit with the optical, radio and soft X-ray data. In
the near feature you can get the result automatically by just
clicking on the estimated onset time that you chose. The second way
is to take into consideration the possibility that particles in
different events might take different path lengths rather than the
nominal 1.2 AU, assuming that particles of different energies were
released simultaneously at the Sun, that the energy of the particle
remains unchanged through the passage from the Sun to $\sim$1 AU,
and that the path length does not depend on the energy. This method
employs the 1/$\beta$ technique discussed in detail by Debrunner,
Fl\"{u}ckiger and Lockwood (1990), Debrunner et al., (1997) and
recently by Huttunen-Heikinmaa et al. (2005). We have added this
calculation in some events that have been analyzed lately. On the
other hand, multi-eruption events can be followed through the
program since we provide the movies related to the time interval for
the intensity-time profile of ERNE.

The well known problem which faces the calculation of the injection
time is the background of the intensity-time profile. In some
channels the background are not reliable to included in the
measurements in both ways. In the interactive program it is easy to
recognize the corrupted background in some energy channels. In Fig.
2 the background of highest channel of the protons intensity seems
not to be in line with the others.

\section{Conclusion}

We have created a program for correlative studies of solar energetic
particles events and we think that this program will be an effective
tool for those kinds of studies according to the following
facts:\begin{enumerate} \item [1-] The program provides a set of
data, concerning the associated phenomenon with each SEP event in
one body of work, which can not be found where else.
\item [2-] The intensity-time profile provided by ERNE instrument
is running simultaneously with the associated phenomenon, so that
the recognition of the association with each change in the
intensity-time profile is easy. \item [3-] The tools for
calculating time and intensity are very easy to use. \item [4-]
The calculation for injection time is available for each event and
for any energy, which will save time and work for researchers.
\item [5-] The previous studies for the events provided by
statistical studies are available on the comments and provide help
in studying the SEP events.
\end{enumerate}
\begin{acknowledgements}
The SOHO is an international cooperation project between ESA and
NASA. The CME catalog is generated and maintained by the Center
for Solar Physics and Space Weather, The Catholic University of
America in cooperation with the Naval Research Laboratory and
NASA.
\end{acknowledgements}

\begin {references}

\reference Cane, H. V., McGuire, R. E., \& von Rosenvinge, T. T.,
ApJ, 301:488-459, 1986.

\reference  Cane, H. V., Reames, D. V., \& von Rosenvinge, T. T.,
JGR, 93, 9555, 1988.

\reference  Chao, J. K., edited by C. T. Russell, pp. 169-174, Inst.
of Geophys. and Planet. Phys., Univ. of Calif., Los Angeles, 1974.

\reference  Chao, J. K., Adv. Space Res., 4, 327-330, 1984.

\reference  Cho, K. S., Kim, K. S., Moon, Y. J., and Dryer, M.,
Solar Phys., 212, 151-163, 2003.

\reference  Cliver, E. W., D. F. Webb, and R. A. Howard, Solar
Phys., 187, 89-114, 1999.

\reference  Debrunner, H., Fl\"{u}ckiger, E. O., and Lockwood, J.
A., APJ supplement series,73,259, 1990.

\reference  Debrunner, H., Lockwood, J. A., Barat, C., et al., APJ,
479, 997, 1997.

\reference  Dodge, J. C., Sol. Phys., 42, 445, 1975.

\reference  Dryer, M., Comments on the origins of coronal mass
ejections, Solar Phys., 169, 421-429, 1996.

\reference  Gopalswamy, N., Kaiser, M. L., Lepping, R. P., Kahler,
S. W., Ogilvie, K.,Berdichevesky, D., Kondo, T., Isobe, T., and
Akioka, M., JGR, 103, 307-316, 1998.

\reference  Gopalswamy, N., M. L. Kaiser, J. Sato, and M. Pick, in
High Energy Solar Physics Workshop-Anticipating HESSI, ASP Conf.
Ser., vol. 206, edited by R. Ramaty and N. Mandzhavidze, p. 351,
Astron. Soc. of the Pacific, San Francisco, Calif., 2000a.

\reference  Gosling, J. T., The solar flare myth, J. Geophys. Res.,
98, 18,937-18,950, 1993.

\reference  Gosling, J. T., and Hundhausen, A. J., Reply, Solar
Phys., 160, 57-60, 1995.

\reference  Huttunen-Heikinmaa K., E. Valtonen, and T. Laitinen,
Proton and helium release times in SEP events observed with
SOHO/ERNE, \textit{A\&A, 442,} 673-685, 2005.

\reference  Kahler, S.W., ApJ, 214, 891, 1977.

\reference  Kahler, S. W.,Astrophysical Journal, Part 1, vol. 261,
p. 710-719, 1982.

\reference  Kahler, S., A\&A, 30, 113-141, 1992.

\reference  Kallenrode, M. B., JGR, 101, No. A11, 24,393-24,409,
1996.

\reference  Klein, K. L., Trottet, G., Space Science Reviews, Volume
95, Issue 1-2, Pages 215 - 225, 2001.

\reference  Klassen, A., H. Aurass, G. Mann, and Thompson, B. J.,
Astron. Astrophys. Suppl. Ser., 141, 357-369, 1997.

\reference  Kocharov, L., Torsti, J., Laitinen, L., \& Teittinen,
M., Solar physics 190: 295-307, 1999.

\reference  Laitinen, T., Klein, K.-L., Kocharov, L., et al., A\&A
360, 729, 2000.

\reference  Leblanc, Y., G. A. Dulk, A. Vourlidas, and Bougeret,
J.-L., J. Geophys. Res., 160, 25,301-25,312, 2001.

\reference Mason, G. M., et al., GRL, 26, 141-144, 1999.

\reference  Moon, Y.-J., G. S. Choe, H. Wang, Y.-D. Park, N.
Gopalswamy, G. Yang, and S. Yashiro, Astrophys. J., 581, 694-702,
2002b.

\reference  Nelson, G. J., \& Melrose, D. B. in Solar Radiophysics,
ed. D. J. McLean \& N. R. Labrum (New York: Cambridge Univ. Press),
333, 1985.

\reference  Neupert, W. M., B. J. Thompson, J. B. Gurman, and S. P.
Plunkett, J. Geophys. Res., 160, 25,215-25,226, 2001.

\reference  Reames, D. V., Space Sci. Rev., 90(3/4), 413-491, 1999.

\reference  Reiner, M.J., M. L. Kaiser, S. P. Plunkett, N. P.
Prestage, and R. Manning, Radio Tracking of a White-Light Coronal
Mass Ejection from Solar Corona to Interplanetary Medium
\textit{ApJ, 529, 1,} L53-L56, 2000.

\reference  Reiner, M. J.,  Kaiser, M. L.,  Gopalswamy, N.,  Aurass,
H., Mann, G.,  Vourlidas, A., and Maksimovic, M., JGR, 106,
25,279-25,290, 2001.

\reference  Shanmugaraju, A., Y.-J. Moon, M. Dryer, and S. Umapathy,
Solar Phys., 215, 164-184, 2003a.

\reference  Sheeley, N. R., Jr.; Bohlin, J. D.; Brueckner, G. E.;
Purcell, J. D.; Scherrer, V. E.; Tousey, R.; Smith, J. B., Jr.;
Speich, D. M.; Tandberg-Hanssen, E.; Wilson, R. M., SoPh, 45,377,
1975.

\reference  Sheeley, N. R., Jr.; Howard, R. A.; Koomen, M. J.;
Michels, D. J., ApJ, 272, 349, 1983a.

\reference  Sheeley, N. R., Jr.; Howard, R. A.; Koomen, M. J.;
Michels, D. J.; Schwenn, R.; Muhlhauser, K. H.; Rosenbauer, H.,
sowi.conf, 693, 1983b.

\reference  Sheeley, N. R., Jr., Howard, R. A., Michels, D. J.,
Koomen, M. J., Schwenn, R., Muehlhaeuser, K. H., Rosenbauer, H.,
Journal of Geophysical Research (ISSN 0148-0227), vol. 90, p.
163-175, 1985.

\reference  Smart, D. F., and M. A. Shea, A simplified model for
timing the arrival of solar flare-initiated shocks, J. Geophys.
Res., 90, 183-190, 1985.

\reference  Svestka, Z., Solar Phys., 160, 53-56, 1995.

\reference Swarup, G., Stone, P. H., \& Maxwell, A. ApJ, 131, 725,
1960.

\reference  Thompson, B. J., J. B. Gurman, W. M. Neupert, J. S.
Newmark, J.-P. Delaboudiniere, O. C. St. Cyr, S. Stezelberger, K. P.
Dere, R. A. Howard, and D. J. Michels, Astrophys. J., 517,
L151-L154, 2000.

\reference  Torsti, J., et al. , Sol. Phys., 162, 505, 1995.

\reference  Torsti, J., Laitinen, L., Vainio, R., Kocharov, L.,
Anttila, A., \& Valtonen, A.,  Solar Physics 175: 771-784, 1997.

\reference  Torsti, J., Kocharov, L., Teittinen, M., Anttila, A.,
Laitinen, L., Mäkelä, P., Riihonen, E., Vainio, R.,  \& Valtonen,
A., J. Geophys. Res., 104, 9903, 1999b.

\reference  Wagner, W. J., and R. M. Astron. Astrophys., 120,
136-138, 1983.

\reference Zhang, J., K. P. Dere, R. A. Howard, M. R. Kundu, and S.
M. White, Astrophys. J., 559, 452-462, 2001.

\end {references}

\end{document}